\documentclass{test3}
\usepackage{graphics}
\usepackage{graphicx}
\usepackage{dcolumn} 
\usepackage{bm}
\usepackage{epsfig}
\usepackage{amsmath}
\usepackage{amsfonts}
\usepackage{amssymb}
\usepackage{times}
\newcommand{\eg}{{\sl e.g.}}
\newcommand{\etal}{{\sl et al.}}

\newcommand{\sto}{SrTiO$_3$} 

\newcommand{\lao}{LaAlO$_3$}
\newcommand{\tith}{Ti$^{3+}$}

\newcommand{\alo}{AlO$ _2 $}
\newcommand{\tio}{TiO$ _2 $}
\newcommand{\muB}{$\mu_{\rm B}$}
\newcommand{\ef}{$E_{\rm F}$}

\begin{document}

\title{Termination control of electronic phases in oxide thin films and interfaces: LaAlO$_3$/SrTiO$_3$(001) }

\author{R. Pentcheva$^1$, R. Arras$^1$,  K. Otte$^1$, V. G. Ruiz$^1$,  and W. E. Pickett$^2$}
\address{$^1$ Department of Earth and Environmental Sciences, Section Crystallography and Center of Nanoscience, University of Munich, Theresienstr. 41, 80333 Munich, Germany \\
$^2$Department of Physics, University of California, Davis, One Shields Avenue, Davis, CA 95616, U.S.A.}


\abstract{
A wealth of intriguing properties emerge in the seemingly simple system composed of the band insulators LaAlO$_3$ and SrTiO$_3$ such as a  two-dimensional electron gas, superconductivity and magnetism.  In this paper we review the current insight obtained from first principles calculations on the mechanisms governing the behavior of thin LaAlO$_3$ films on SrTiO$_3$(001). In particular, we explore the strong dependence of the electronic properties on the surface and interface termination, the finite film thickness, lattice polarization and defects. A further aspect that is addressed is  how the electronic behavior and functionality  can be tuned by a SrTiO$_3$ capping layer, adsorbates and metallic contacts. Lastly, we discuss recent reports on the coexistence of magnetism and superconductivity in this system for what they might imply about the electronic structure of this system.}

\maketitle
\section{Introduction}
The advance of thin film growth techniques like pulsed laser deposition (PLD) and molecular beam epitaxy (MBE) allows the fabrication of single terminated oxide interfaces on the atomic scale. Understanding the novel phenomena arising in these artificial materials is  not only of fundamental interest, but is also relevant for the development of future electronics and spintronics devices. A system that has attracted most of the interest so far is comprised of the simple band insulators LaAlO$_3$ (LAO) and SrTiO$_3$ (STO) where a two-dimensional electron  gas (2DEG)~\cite{Ohtomo:Hwang2004}, superconductivity~\cite{Reyren:Thiel:Mannhart2007}, magnetism~\cite{Brinkman:Huijben:etal2007} and even signatures of their coexistence~\cite{Bert:2011,Dikin:2011,Li:2011} have been reported. A further feature of interest and potential importance is that the electronic properties can be tuned by the LAO thickness and the system undergoes a transition from insulating to conducting behavior at around four monolayers (ML) LAO~\cite{Thiel:Hammerl:Mannhart2006}. This insulator-metal transition (IMT) can be controlled reversibly via an electric field, e.g. by an atomic force microscope (AFM) tip~\cite{Cen:Thiel:Hammerl:etal2008,Bi:Levy:2010,Chen:2010} and several electronic devices based on this feature have been proposed~\cite{Cen:2009,Bogorin:2010}. The electronic properties can be further tuned by an additional STO capping layer that triggers the IMT already at 2ML thereby stabilizing an electron-hole bilayer~\cite{PentchevaPRL:2010}. The electronic properties show a strong dependence on the growth conditions: varying the oxygen partial pressure in the PLD chamber from 10$^{-6}$ to 10$^{-3}$ mbar induces an increase in sheet resistance by seven orders of magnitude.\cite{Brinkman:Huijben:etal2007} This enormous change implies that oxygen defects play a controlling role during low pressure growth.

Extensive theoretical and experimental efforts aim at explaining the origin of these interfacial phenomena. A central feature is  the polar discontinuity that emerges at the interface: in the (001) direction  LAO consists of charged (LaO)$ ^ {+} $  and  (\alo)$ ^{-} $ planes, while in STO formally neutral (SrO)$ ^ {0} $ and (\tio)$ ^{0} $ planes alternate.  Because both cations change across the interface, two distinct interfaces can be realized:  an electron doped $n$-type with a LaO layer next to a \tio\ layer and a hole-doped $p$-type interface with a SrO and \alo\ next to each other. Assuming formal ionic charges, the electrostatic potential produced by an infinite stack of charged planes would grow without bounds, a phenomenon known as the {\sl polar catastrophe}~\cite{Hwang2006,Goniakowski:2008}. This polar catastrophe cannot of course actually occur, and it may be avoided in several ways: A mechanism also common to semiconductors is an {\sl atomic reconstruction} via introduction of defects or adsorbates. In transition metal oxides  electronic degrees of freedom may lead to an alternative compensation mechanism, i.e. an {\sl electronic reconstruction}.  The latter can give rise to exotic electronic states, for example the observed 2D conductivity in LAO/STO, charge/spin/orbital order,  excitonic or  superconducting phases. A further question that arises is whether a thin film of LAO with only a few layers approaches the regime of the  {\sl polar catastrophe}.  Sorting out the many possibilities is  challenging due to the strong dependence on growth method and conditions.

In this paper we review the progress made so far in understanding the mechanisms that determine the electronic behavior  of LAO/STO(001) based on density functional theory calculations. For furtehr  reviews on the experimental and theoretical work the reader is referred to Refs. \cite{Pauli:2008,Huijben:2009,Pentcheva:2010,Chen:Kolpak:Ismail2010,Triscone:2011}.
In particular we address  finite size effects  and the role of electrostatic boundary conditions~\cite{Stengel:2011} in thin LAO films on STO(001). We first discuss intrinsic mechanisms that arise at defect-free $n$- and $p$-type interfaces providing a possible explanation for  the thickness dependent IMT. Moreover, we explore  the effect of the surface termination on the band diagram. In Sections \ref{STOcap} and \ref{metallic_contact} we address the role of a STO capping layer and a metallic electrode. A controversially discussed issue is the presence of an internal potential buildup within the LAO film, as predicted by DFT calculations on LAO/STO(001) with abrupt interfaces. While recent AFM experiments provide evidence for such an internal field in terms of a polarity-dependent asymmetry of the signal~\cite{Xie:Hwang:2010}, x--ray photoemission studies~\cite{Segal:Reiner:etal2009,Sing:Claessen:2009,Chambers:Ramasse:2010} have not been able to detect shifts or 
 broadening of core-level spectra that would reflect an internal electric field. 
This discrepancy implies that  besides the electronic reconstruction, extrinsic effects play a role, e.g. oxygen defects~\cite{Zhong:2010,Bristowe:2011}, adsorbates such as  water or hydrogen~\cite{Son:2010} or cation disorder~\cite{Willmott:2007,Qiao:2010}. An overview of the current theoretical understanding of the role of such extrinsic mechanisms is provided in Section~\ref{sec:defects}. Finally, in Section~\ref{Broader_Functionalities} we address recent reports on the coexistence of  superconductivity and ferromagnetism in LAO/STO - a phenomenon which may be helpful in understanding the underlying electronic structure in these nanoscale systems.

\section{\label{Polar_oxide}Polar oxide films on a nonpolar substrate: LAO/STO(001)}

\begin{figure}[ht]\vspace{-0pt}
   \begin{center}
\includegraphics[scale=0.35]{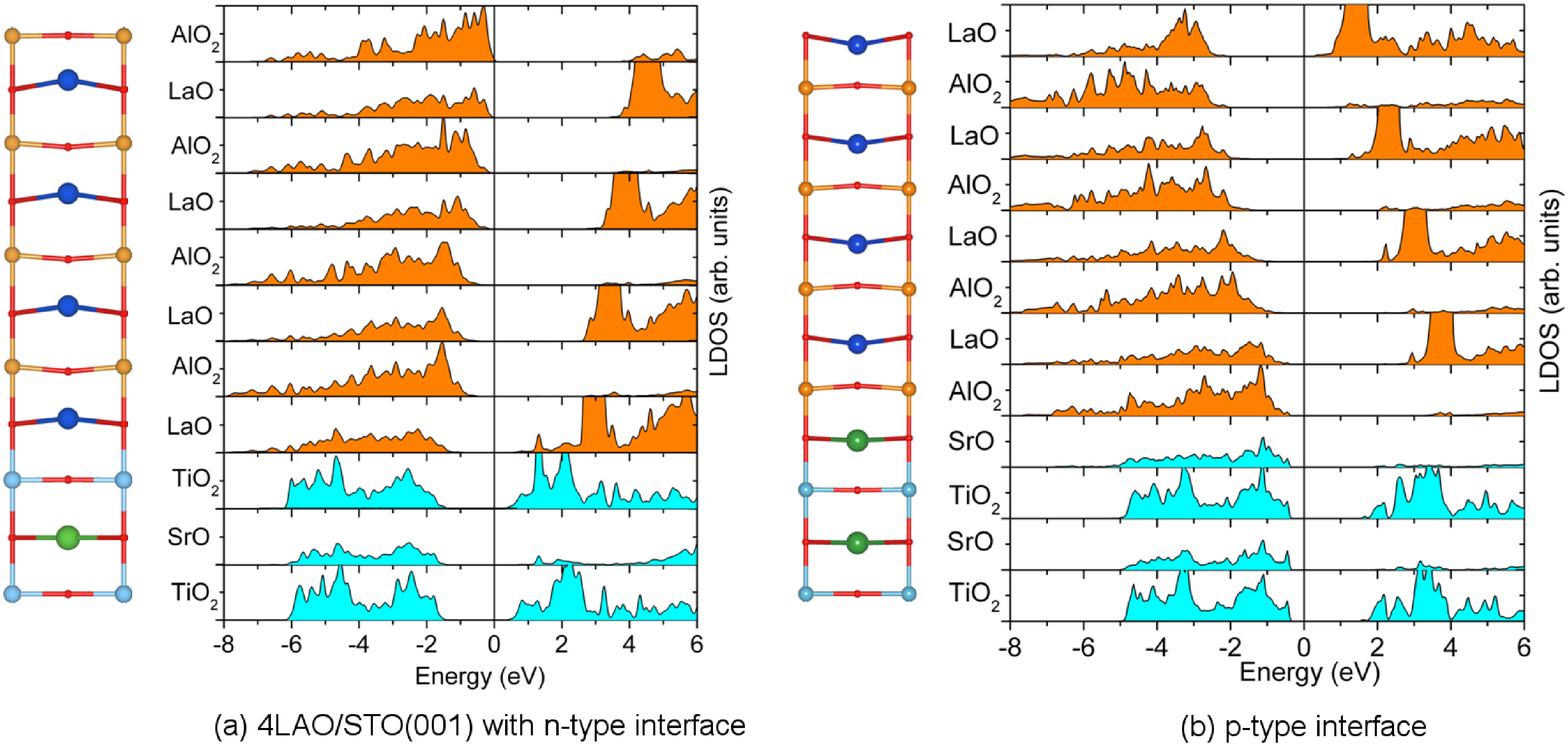}
   \end{center}
\caption{\label{fig:4laonp} (Color online) Side view of the relaxed structure and layer resolved DOS of 4LAO/STO(001) with a (a) $n$-type  and (b) $p$-type interface. Note the emergence of a potential build-up and a lattice polarization within the LAO film of opposite sign for the $n$ and $p$-type interface.}
\end{figure}

\begin{figure}[t]\vspace{-0pt}
   \begin{center}
\includegraphics[scale=0.3]{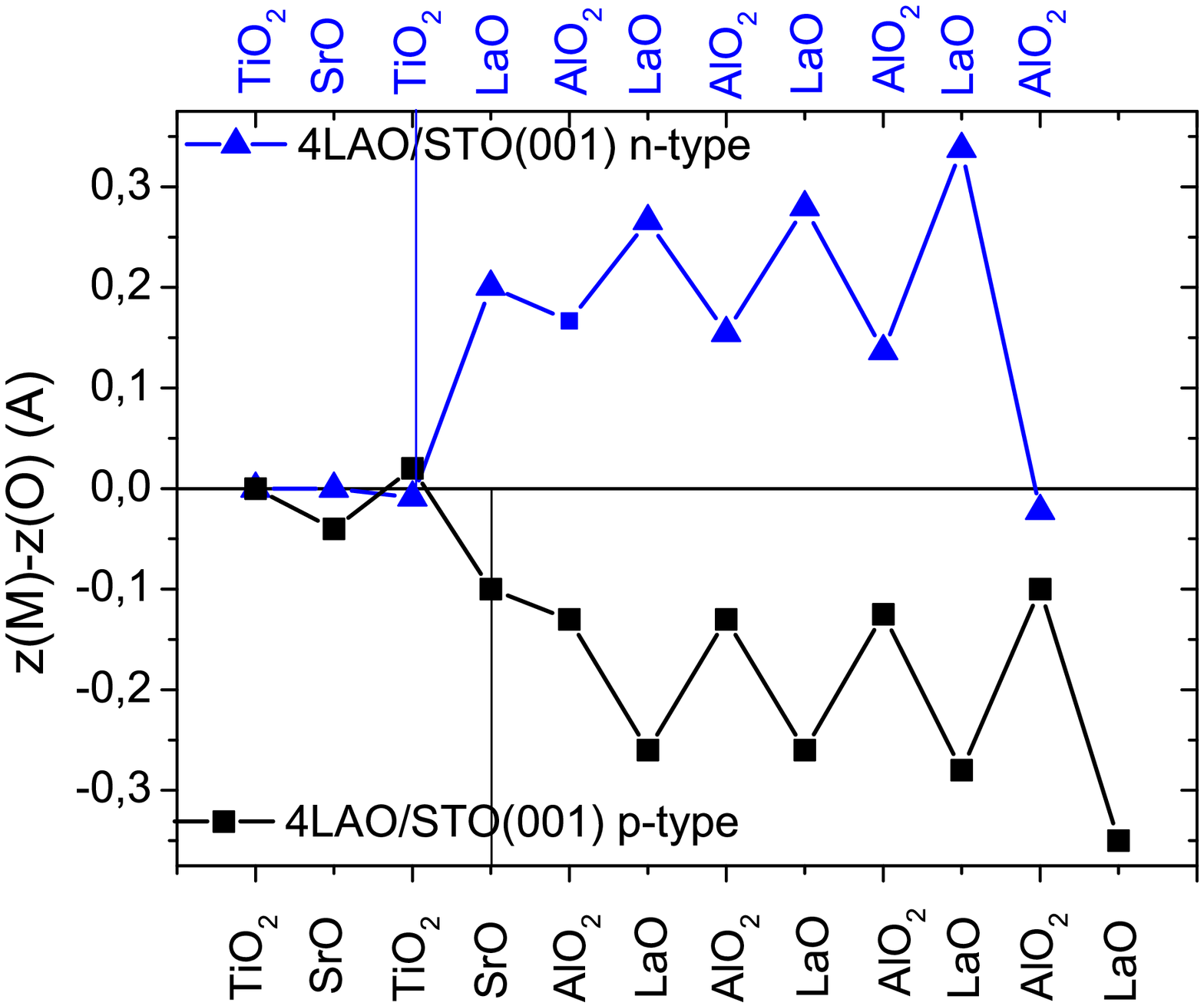}
   \end{center}
\caption{\label{fig:zM-zOcapped} (Color online) Oxygen-cation buckling in 4LAO/STO(001) with an $n$- (blue triangles) and a $p$-type interface (black squares). As already shown in the side view in Fig.~\ref{fig:4laonp}, the lattice polarization has an opposite sign for the $n$. and $p$-type interface.}
\end{figure}

An intriguing experimental finding in LAO/STO(001) is the thickness dependent IMT in thin LAO films on STO(001)~\cite{Thiel:Hammerl:Mannhart2006}. 
Density functional theory calculations (DFT)  demonstrate that an internal electric field emerges in  thin polar LAO overlayers on STO(001)~\cite{Ishibashi:Terakura2008,Pentcheva:Pickett2009,Son:2009,Pentcheva:2010}. As shown in the layer resolved density of states (DOS) of a 4ML LAO film on STO(001) with an $n$-type interface (Fig.~\ref{fig:4laonp}a), this is expressed in an upward shift of the O $2p$ bands as they get closer to the surface. Interestingly, a large electric field of opposite sign arises in an analogous manner in a defect free 4ML LAO/STO(001) overlayer with a $p$-type interface (Fig.~\ref{fig:4laonp}b). The internal electric field of the polar LAO film is screened to a large degree (but not completely) by a strong lattice polarization. As can be seen from the side view of the relaxed structure  in Fig.~\ref{fig:4laonp}a and the anion-cation buckling displayed in Fig.~\ref{fig:zM-zOcapped},  for an $n$-type interface this lattice polarization  is characterized by a strong outward shift of 
 La by 0.2-0.3\AA\ and buckling in the subsurface \alo\ layers while the surface layer shows similar relaxations for anions and cations. Experimental evidence for an internal electric field   within the LAO film seems contradictory.  Segal {\it et al.}\cite{Segal:Reiner:etal2009} and Chambers {\it et al.}\cite{Chambers:Ramasse:2010} found no evidence for the expected core level shifts due to an internal field.  However, evidence for a lattice polarization as a respoce to an internal electric field has been obtained  from surface x-ray diffraction (SXRD) by Pauli \etal~\cite{Pauli:2011}.  The latter study detected also a dependence of the lattice response on the LAO thickness with a maximum buckling in a 2 ML LAO film. This indicates possibly  a stronger role of cation interdiffusion or higher affinity for adsorbates in the thicker films. Such extrinsic mechanisms can reduce the internal potential and thereby the  lattice polarization (see discussion in Section~\ref{sec:defects}). 

The lattice relaxation has a crucial effect on the electronic properties: if the atoms are fixed at their ideal bulk positions, all systems are metallic~\cite{Pentcheva:Pickett2009}. The  lattice polarization in the relaxed system allows several layers of LAO to remain insulating with a band gap of 1.7~eV for 1ML LAO/STO(001) and a gradual decrease by $\sim0.4$~eV per added LAO ML. Finally, at  around  5 monolayers of LAO an electronic reconstruction takes place. However, the closing of the band gap is {\sl indirect} in real space as it is due to overlap of the valence band defined by the O $2p$ band in the surface layer and the conduction band marked by Ti $3d$ states at the interface. Consequently, the carrier density is much lower than the one expected if  $0.5e$ were transferred to the interface. Furthermore, the results suggest carriers of two different types: electrons at the interface and holes in the surface layer. We will return to the possibility of excitonic effects later when discussing the role of a 
 STO capping layer in Section~\ref{STOcap}.

For a defect-free 4LAO/STO(001) with a $p$-type interface the ionic relaxations are of similar magnitude but of opposite sign (note the inward relaxation of the La-ions and the anion cation buckling shown in Fig.~\ref{fig:zM-zOcapped}). A very similar thickness dependent insulator-to-metal transition is expected which involves however different states:  La $5d$ states in the surface LaO layer and  O $2p$ states at the interface. In experiments, the $p$-type interface has been found so far insulating~\cite{Ohtomo:Hwang2004}  and has therefore attracted much less attention. The analysis of O K-edge spectra by Nakagawa, Hwang, and Muller suggested that compensation takes place via oxygen vacancies,~\cite{Nakagawa:2006} that is, by atomic reconstruction. Still, the results above and by~\cite{Ishibashi:Terakura2008} show that provided defect-free LAO/STO interfaces with a $p$-type interface can be realized they would exhibit just as interesting thickness dependent properties as the ones for the $n$-type interface.  



So far we have considered stoichiometric 4ML LAO films on STO(001), where e.g. for an $n$-type interface the LAO film is terminated by an AlO$_2$ surface layer. 
Pavlenko and Kopp~\cite{Pavlenko:2011} recently investigated LAO/STO(001)  with an $n$-type interface and a LaO surface termination. Note that here both the surface and the interface are electron doped. Not surprisingly, the system is found to be metallic for 2.5 and 3.5 ML LAO with finite occupation of Ti $3d_{xy}$ states in the interface layer and La $5d_{x^2-y^2}$ states in the surface layer. The authors interpreted the surface La $5d$ occupation as a result of surface tensile stress. While the latter influences e.g. the vertical relaxation of ions,  the polarity discontinuity at the surface and interface is likely to be the dominating effect.

\section{\label{STOcap}Influence of a STO capping layer}

\begin{figure}[t!]

   \begin{center}  
\includegraphics[scale=0.43]{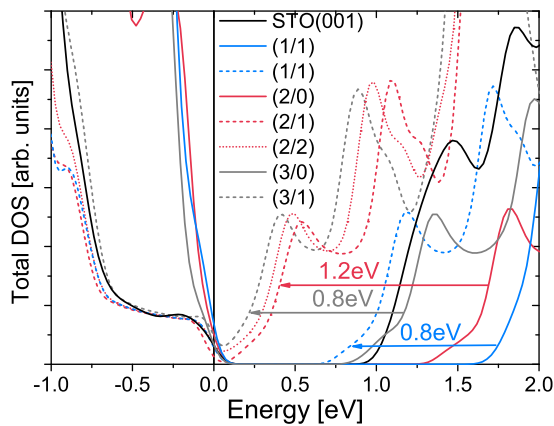}
   \end{center}
\caption{(Color online) Total density of states (LDOS) $n$LAO/STO(001), capped by $m$ STO layers. Note that the band gap closes for $n>2$ML as soon as a capping STO layer is added. Thereby, the effect of the first STO layer is most pronounced.  }\label{fig:tdosmSTOnLAOSTO} 
\end{figure}
In  contrast to the sharp IMT in LAO films on STO, Huijben \etal~\cite{Huijben:2006} observed a much smoother transition from insulating to conducting behavior that starts already at 2ML LAO if this latter is covered by a nonpolar STO layer. The total DOS of  $n$ ML LAO/STO(001) covered by $m$ layers of STO [denoted in the following as $(n/m)$] is displayed in Fig.~\ref{fig:tdosmSTOnLAOSTO}. The DFT results reveal that for $n=2$ ML already a single STO capping layer leads to an IMT. Increasing the number of STO capping layers $m$ or LAO layers $n$ enhances the DOS at the Fermi level, but the first STO  capping layer has the most dramatic effect, reducing the band gap by $\sim 1.0$~eV. 

\begin{figure}[t!]
   \begin{center}
\includegraphics[scale=0.43]{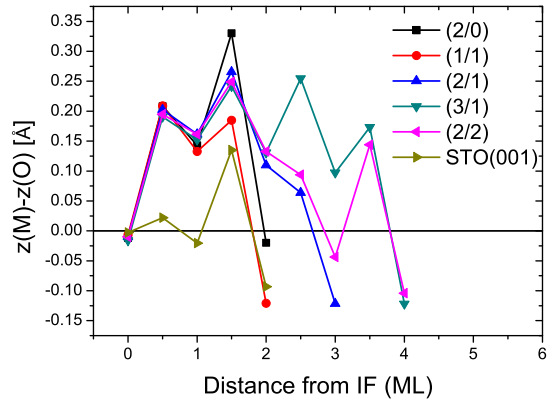}
   \end{center}
\caption{(Color online) Cation-anion vertical buckling in $m$STO/$n$LAO/STO(001) and at the STO(001) surface for comparison.  }\label{fig:structcapped} 
\end{figure}
In order to gain more insight into the mechanism of closing of the gap, we analyze first the lattice relaxations. Fig.~\ref{fig:structcapped} shows the cation-anion displacements in $m$STO/$n$LAO/STO(001). We observe slight reduction in the buckling within the LAO film compared to the (2/0) system  and a polarization within the STO layer.  The relaxation pattern in the STO capping layer is in fact  similar to the one of the STO(001) surface~\cite{Padilla1998} (for comparison the relaxations at a STO(001) surface are  added to Fig.~\ref{fig:structcapped}). However, the total contribution of the STO capping layer is small as the ionic dipole moments  of the SrO and TiO$_2$ layers are of opposite sign (negative dipole moment in the surface TiO$_2$ layer and a positive one in the subsurface SrO layer) and nearly cancel. The results show that the ionic contribution to the dipole moment in $m$STO/$n$LAO/STO(001) scales with the number $n$ of LAO layers and increases roughly by 1~e\AA\ per added LAO layer.

\begin{figure}[t!]
   \begin{center}
\includegraphics[scale=0.33]{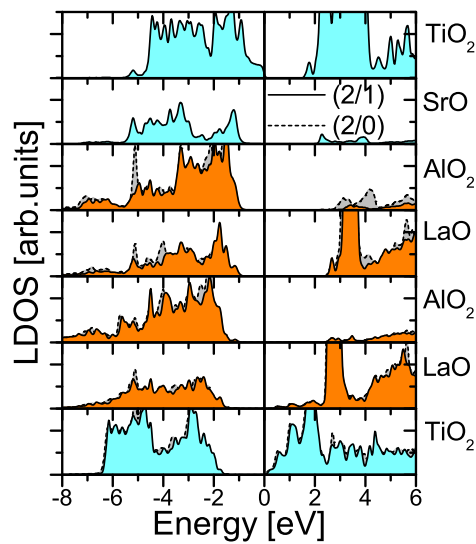}
   \end{center}
\caption{(Color online) Layer resolved density of states (LDOS) of 2LAO/STO(001) with (black line) and without (dashed line) a single STO capping layer. A dispersive O $2p$ band in the surface TiO$_2$ layer leads to a closing of the band gap in the capped system.\cite{PentchevaPRL:2010}  }\label{fig:dosmSTOnLAOSTO} 
\end{figure}
The impact of the STO capping layer turns out to be of electronic origin: Fig.~\ref{fig:dosmSTOnLAOSTO} shows the layer resolved DOS of (2/0) (dashed line) and (2/1) (solid line) line aligned at the bottom of the Ti $3d$ band at the interface. Both in the capped and uncapped system  the O~$2p$ bands  within the LAO film exhibit a gradual upward shift of 0.4~eV per LAO as they approach the surface. In the capped system, there is an additional strong shift/broadening of the O~$2p$ band in the surface TiO$_2$ layer of $\sim 0.8$~eV that closes the band gap and induces an electronic reconstruction~\cite{PentchevaPRL:2010}.

\begin{figure}[t!]
   \begin{center}
\includegraphics[scale=0.43]{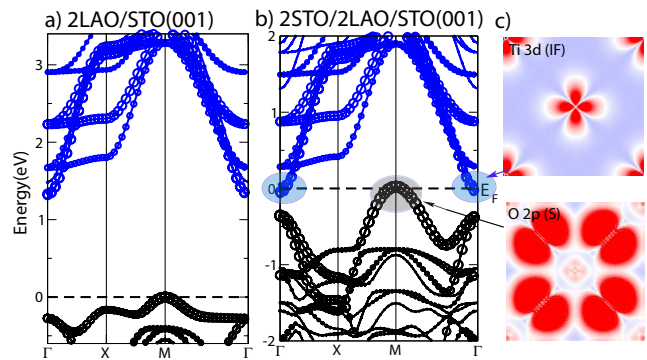}
   \end{center}
\caption{(Color online) Band structure of a) 2LAO/STO(001) with an indirect band gap and b) 2STO/2LAO/STO(001) with a closed band gap due to overlap of an electron band at  $\Gamma$ and a holes band at $M$; c) integrated electron density around the Fermi level shows electrons of Ti $3d_{xy}$  character at  $\Gamma$ and holes of O~$2p_{\pi}$ type at  $M$. Adapted from Ref.\cite{PentchevaPRL:2010}.  }\label{fig:bandmSTOnLAOSTO} 
\end{figure}
Indeed, the band structure in Fig.~\ref{fig:bandmSTOnLAOSTO} shows  a dispersive O~$2p$ surface band, similar to a surface state in STO(001)~\cite{Padilla1998, Kimura1995}, that extends 0.8~eV above the subsurface O~$2p$ band and marks the top of the valence band at the $M$ point of the Brillouin zone. On the other hand the bottom of the conduction band lies at the $\Gamma$ point and is determined by Ti $3d$ states in the interface layer. Thus the closing of the band gap is indirect both in real and reciprocal space. Increasing the number of LAO and STO capping layers enhances the overlap of the valence and conduction bands at the Fermi level.

The electron density distribution in 2STO/2LAO/STO(001) integrated between -0.3 and 0.0~eV shows electrons of Ti~$3d_{xy}$ character in the interface layer and holes in the O~$2p_{\pi}$ bands at the surface.  From the curvature of the electron and hole bands at the $\Gamma$ and $M$ points we determine a significantly higher effective mass of the holes (1.4~m$_e$/1.2~m$_e$ in the uncapped/capped system, respectively) than for the  lighter electrons (0.4~m$_e$). The presence of two types of carriers with different mobilities is confirmed in Hall and magnetoresistance measurements~\cite{PentchevaPRL:2010} where the data between 0-100~K can be fitted only by using a two-band model with one hole and one electron band. Photoemission experiments give further evidence for the presence of holes in the surface layer~\cite{PentchevaPRL:2010}. Thus the presence of a nonpolar oxide capping layer seems to stabilize the system with respect to surface defects and adsorbates that are likely to eliminate holes in the uncapped systems.

\begin{figure}[t!]
   \begin{center}
\includegraphics[scale=0.3]{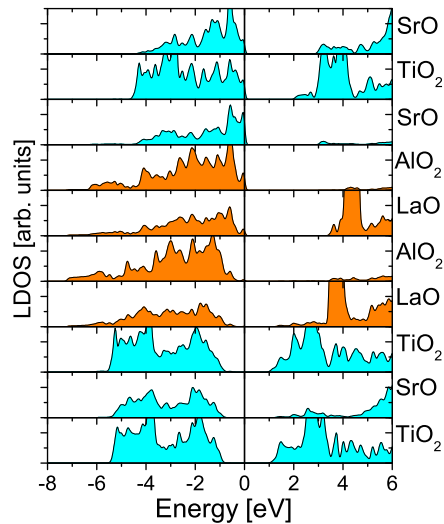}
   \end{center}
\caption{(Color online) Layer resolved density of states (LDOS) of a SrO terminated 1.5ML STO/2LAO/STO(001), showing insulating behavior.  }\label{fig:dos15STO2LAOSTO} 
\end{figure}
In contrast, when the system is terminated by a SrO layer (using a 1.5 ML STO overlayer) the layer resolved DOS in Fig.~\ref{fig:dos15STO2LAOSTO} shows that the band gap of $\sim 1$ eV of the uncapped (2/0) system is preserved. The reduction of the band gap is solely due to the potential build up within the LAO film, while the valence band (VB) of the STO capping layer aligns with the VB in the top AlO$_2$ layer without the surface state characteristic of the TiO$_2$-terminated capping layer. This system exhibits a behavior that is closer to what one would expect when adding a ``nonpolar'' oxide overlayer on top of LAO/STO(001).

\section{\label{sec:defects}Role of lattice defects and adsorbates}

{\sl Cationic intermixing} The systems considered so far contained abrupt interfaces. However, several studies suggest significant  cation intermixing at the interface~\cite{Willmott:2007,Qiao:2010,Pauli:2011}. Intermixing has been discussed as an alternative mechanism to the electronic reconstruction to compensate polarity, but the role of intermixing in canceling the potential divergence is questioned~\cite{Hwangcondmat}. Qiao \etal~\cite{Qiao:2010} proposed as main sources of defects deviations from a 1:1 La:Sr ratio due to strong angular dependence during the PLD process, Sr vacancy formation, and La interdiffusion into the STO substrate. They investigated La$\rightarrow$Sr and  Al$\rightarrow$Ti substitution both near and far from the interface using up to 4 ML LAO on a 6 ML thick STO substrate in a $c(2\times 2)$ lateral unit cell.
The DFT results indicate that coupled La-Sr, Al-Ti exchange processes involving both the surface and interface region lower the energy of  the system and the energy gain is dependent on the LAO thickness, being  $\sim 1.0$~eV for $n=3$ LAO and  $\sim 1.6$~eV for $n=4$ LAO layers, respectively. A comparison of the calculated and measured valence band offset was used as a further criterion to prove the presence of intermixing. 

Pauli \etal~\cite{Pauli:2011} performed SXRD experiments and DFT calculations on films with thickness between 2-5 ML LAO. The experimental data indicated an 80\% filling of the surface layer with 20\% of a layer on top of that.  Futhermore the results were consistent with cation intermixing exceeding 5\% thoughout 3 ML around the interface layer. In contrast to the study of Qiao \etal~\cite{Qiao:2010}, DFT calculations   on abrupt and intermixed interfaces did not show significant differences in electronic behavior.

{\sl Oxygen defects} As mentioned in the introduction, a number of experimental studies have  shown the strong influence of the oxygen partial pressure during deposition or post-deposition annealing on the transport properties. Despite its importance, there are only a few studies that have addressed the role of oxygen vacancies. Cen \etal~\cite{Cen:Thiel:Hammerl:etal2008} considered a 3ML LAO/STO(001) with a vacancy in the surface AlO$_2$ layer and found a dramatic difference of the electronic properties with reduction/cancellation of the electric field within the LAO film and an accumulation of carriers at the interface, in contrast to the insulating  defect-free 3 ML LAO/STO(001). Chen, Kolpak and Ismail-Beiji~\cite{ChenPRB:2009} modeled LAO/STO superlattices with the $p$-type interface and observed that oxygen vacancies are repelled from the interface towards bulk STO.    On the other hand, Zhong, Xu and Kelly~\cite{Zhong:2010} studied oxygen vacancies in an (LAO)$_m$/(STO)$_m$ superlattice with alternating  $n$- and $p$-type interfaces. In contrast to Chen \etal~\cite{ChenPRB:2009}, they found that the formation energy of vacancies is lowest at the $p$-type interface and  $\sim 2$ eV higher at the $n$-type interface with a nonlinear dependence in between. Furthermore, the vacancy formation tends to be more favorable in BO$_2$-layers than in AO-layers of the perovskite lattice. The formation energy at the $p$-type interface becomes negative with respect to the one in STO bulk at a critical thickness $m=3-4$ MLs. The electrons released by the vacancy are transferred to the conduction band minimum at the $n$-type interface (with a significant orbital polarization in the interface and more distant TiO$_2$ layers) while the $p$-type interface is insulating. This result implies separation between regions where impurity scattering takes place and regions of enhanced carrier density. The  authors also observed that the vacancy formation reduces the core-level shifts within the superlattice and proposed this as a possible explanation of the  lacking broadening in  XPS measurements~\cite{Segal:Reiner:etal2009,Sing:Claessen:2009,Chambers:Ramasse:2010}.

Zhang \etal~\cite{Zhang:2010} studied the formation of vacancies in thin LAO overlayers on STO(001) using an asymmetric setup with dipole correction, and varied the position of the vacancies within the LAO film. Note that only the $\Gamma$-point was used for integration within the Brillouin zone. In systems with a $p$-type interface  vacancies are formed preferentially at the interface, thereby compensating the valence discontinuity and leaving the system insulating. In contrast, for systems with an $n$-type interface the formation energy was significantly higher. Furthermore, vacancies tend to be formed rather in the surface LAO layer than at the interface.  This has two consequences: it compensates the $p$-type surface AlO$_2$-layer and enhances the carrier density at the interface. Based on a phenomenological electrostatic model, Bristowe, Littlewood and Artacho~\cite{Bristowe:2011}  showed that there is a critical thickness for surface vacancy formation, in agreement with DFT calculations~\cite{
 Li:2009}. Furthermore, they proposed that surface defects generate a  trapping potential for carriers whose strength depends on the  LAO-film thickness. 

DFT calculations~\cite{pavlenko2011} introducing an oxygen vacancy in the interface TiO$_2$ layer within a $2\times1$ lateral unit cell indicate spin-polarization of the carriers in the Ti $3d$ band and propose this as a possible explanation of the recently reported coexistence of ferromagnetism and superconductivity in the LAO/STO(001) system (see also discussion in Section~\ref{Broader_Functionalities}).  Further studies are necessary to explore lower concentrations of vacancies and other thermodynamically more stable sites for the vacancy (\eg{} in the surface AlO$_2$-layer).

{\sl H-adsorption} In most of the experiments so far, the samples are exposed to air, thus the interfacial properties can be significantly affected by adsorbates. These are considered as a  possible origin of  reversible writing and erasing of conducting regions on LAO/STO(001) by an AFM tip~\cite{Cen:Thiel:Hammerl:etal2008}. Son et al~\cite{Son:2009} investigated the H-adsorption using DFT and the generalized gradient approximation (GGA)~\cite{GGA}. They found that the adsorption energy increases with LAO thickness and that beyond a critical thickness of $\sim 5$ ML the energy gain exceeds the energy needed for H$_2$ or H$_2$O dissociative adsorption. An interesting feature is  the planar adsorption geometry with the  H-O bond being nearly parallel to the surface reflecting a maximized overlap between the H $1s$ orbitals and the planar O $2p_{\pi}$ states. H is found to donate electrons to the STO conduction band. Furthermore, size and sign of the internal potential buildup can be tuned by  the H concentration on the surface: e.g. the latter 
 is exactly canceled for a $(2\times 1)$ surface unit cell (this coverage corresponds to $0.5e$ transfer per $(1\times 1)$ lateral unit cell). The impact of H-adsorption bears strong parallels to the influence of a metallic contact layer on the interfacial properties that is discussed in the next Section. The influence of further adsorbates like water has not been considered so far but needs to be addressed in order to obtain a comprehensive picture of all effects that may determine the properties of LAO/STO(001).

\section{\label{metallic_contact}Impact of metallic contacts on the electronic behavior of LAO/STO(001)}

In Sections~\ref{STOcap} and \ref{sec:defects} we have seen that the presence of a nonpolar oxide overlayer  or defects can significantly alter the electronic properties of the LAO/STO system. For device applications metallic overlayers are important and several experimental studies have used setups with metallic contacts on LAO/STO(001)~\cite{Chen:2010, Jany:2010, Bhalla:2011}. The coupling between metallic thin films  and semiconductors or ferroelectric oxides have been widely studied ~\cite{Campbell:1997, Goniakowski:2004, Duan:2006, Fu:2007, Fechner:2008}. In order to gain understanding in the impact of metallic contacts on the electronic behaviour of an oxide interface, we have recently performed DFT calculations on a series of metal electrodes on LAO/STO(001)~\cite{Ruiz2011}. Here we discuss how the electronic properties can be influenced by the metallic contact. 

\begin{figure}[t]
 \hspace{-2cm}
   \begin{center}
   \includegraphics[scale=1.2,angle=0]{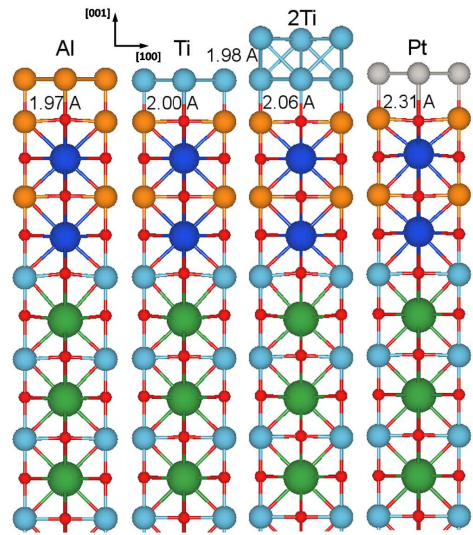}
   \end{center}
\caption{ Side view of the relaxed structures for M/LAO/STO systems (M = Al, Ti, 2Ti and Pt). Only half of our symmetric slab is shown. M-O bond lengths near the surface increase from Al to Pt. In contrast, the buckling in the interfae TiO$_2$ layer decreases. } \label{fig:structAlTiPt} 
\end{figure}  

All the calculations presented in this Section have been performed with the full potential linearized augmented plane wave method in the WIEN2K implementation~\cite{Wien2k} using  GGA~\cite{GGA}. We have also explored the influence of an on-site Coulomb repulsion parameter (LDA/GGA+$U$ method~\cite{Anisimov:1993}) with  $U = 5$ eV and $J=1$~eV for the Ti $3d$ orbitals, and $U=7$ eV for the La $4f$ orbitals. The irreducible part of the  Brillouin zone was sampled with 21 k-points. The systems were modeled by a symmetric slab, with a 2 and 4 ML LAO film on a TiO$_2$-terminated STO substrate with a lateral lattice parameter set to the GGA-value of bulk STO (3.92 \AA). The influence of the substrate thickness was probed by using a system with a thin (2.5 ML) and thick (6.5) ML STO slab. The metal atoms were adsorbed on top of the oxygen atoms in the top AlO$_2$ layer. These sites were found to be energetically favorable in
  previous studies of metallic overlayers on LAO(001)~\cite{Asthagiri:Sholl:2006} and  for other transition metal / perovskite oxide interfaces~\cite{Asthagiri:Sholl:2002,Ochs:2001,Oleinik:2001}. Further metallic layers are deposited assuming a fcc stacking of the layers. A vacuum of at least 10 \AA{} was used to separate the supercell from its periodic images and avoid spurious interactions. The atomic positions were relaxed within tetragonal symmetry. A side view of the relaxed systems with an Al (1 ML), Ti (1 and 2 ML) and Pt (1 ML) overlayer is  shown in Fig.~\ref{fig:structAlTiPt}.

\subsection{Impact of the contact layer on the electronic properties}
\begin{figure}[t]
 \hspace{-2cm}
   \begin{center}
   \includegraphics[scale=0.7,angle=0]{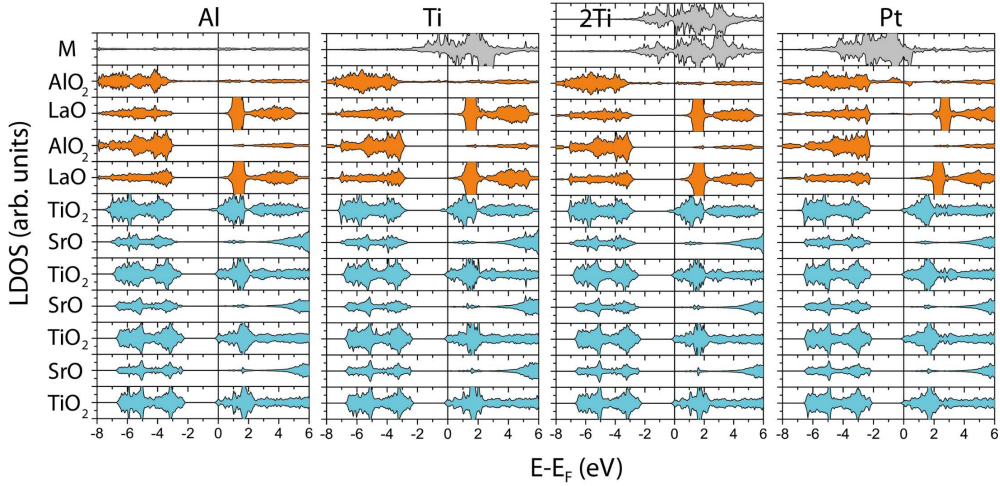}
   \end{center}
\caption{ Layer resolved density of states for 2 ML LAO/STO(001) with a Al (1ML), Ti (1 and 2 ML) and Pt (1ML). The internal electric field, observed in the uncovered films, is cancelled for an Al and Ti contact. A residual field is obtained for a Pt electrode. } \label{fig:DOSAlTiPt2LAOSTO} 
\end{figure}  

As discussed in Section \ref{Polar_oxide}, in the $n$LAO/STO(001) system a central  feature  is the upward shift of the O $2p$ bands within the polar LAO film.  The layer resolved DOS of Al, Ti and Pt contact layers on top of 2ML LAO/STO(001) are shown in Fig.~\ref{fig:DOSAlTiPt2LAOSTO}. Upon adsorption of an Al or Ti overlayer this potential build up is largely canceled. In contrast, for Pt there is a nonvanishing shift of the unoccupied La  $4f$ and $5d$ bands. In all cases, both the surface contact layer and the interface (IF) layer are metallic with a significant occupation of the Ti $3d$ band at the interface and a decreasing occupation in deeper layers.  
\begin{figure}[t]
 \hspace{-2cm}
   \begin{center}
   \includegraphics[scale=0.17,angle=0]{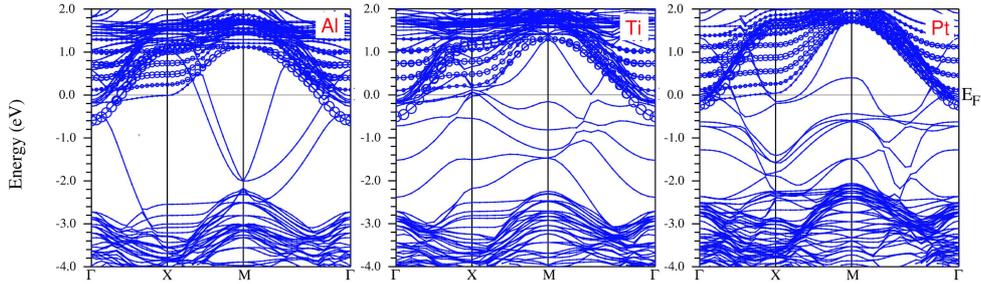}
   \end{center}
\caption{ Majority spin band structures for $M$/LAO/STO systems ($M$ = Al, Ti and Pt). The contribution of Ti-$3d$ states from  the interface TiO$_2$ layer is marked by circles.} \label{fig:bandsAlTiPt} 
\end{figure}

As can be seen from the band structure in Fig.~\ref{fig:bandsAlTiPt}, multiple bands contribute to conductivity: the lowest lying bands at the interface are of Ti $3d_{xy}$ character, in the IF-1 layer all $t_{2g}$ states contribute, while in deeper layers $d_{xz}$, $d_{yz}$ bands lie lowest in energy. The $d_{xy}$ bands have a strong dispersion along $M-\Gamma -X$, whereas $d_{xz}$ and $d_{yz}$ are heavier along $\Gamma -X$. This difference in velocities (masses) suggests different mobilities of the carriers. Further bands between \ef\ and \ef-2.0~eV stem from the surface metallic layer. We observe that the occupation of Ti $3d$ bands at the interface depends strongly on the type of metal contact on the surface: for an Al, Ti and Pt overlayer the band edge is 0.65, 0.5 and 0.25~eV below the Fermi level at $\Gamma$.

 \begin{figure}[t]
\hspace{2cm}
   \includegraphics[scale=0.4,angle=0]{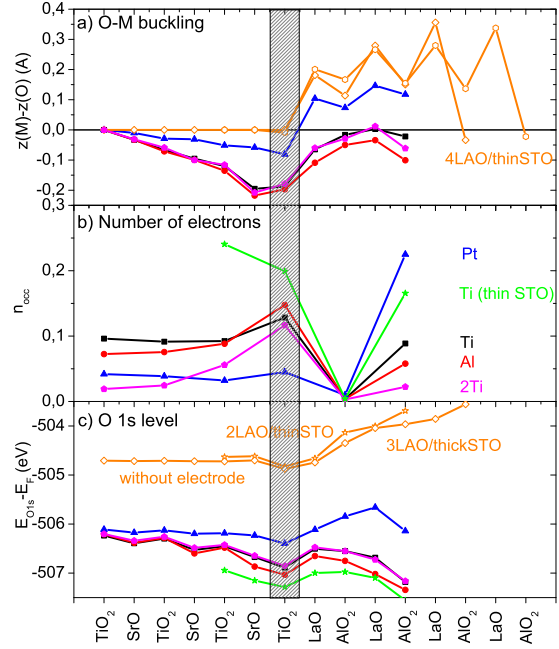}
\caption{ a) Oxygen-cation buckling along the [001] direction. b) Layer resolved electron occupation integrated between $E_F-0.65$ eV and $E_F$. Charge on the LaO and SrO layers is zero (not shown). c) Positions of O$1s$ states with respect to $E_F$. Data for 1ML Ti and thin /thick STO substrate is marked by green stars/black squares; 2ML Ti (magenta pentagons), as well as 1ML Pt (blue triangles) and 1ML Al (red circles). Results for $n$LAO/STO(001) without a metallic overlayer are displayed  in orange with empty symbols (stars, diamonds, and hexagons for  $n=2,3,4$ ML, respectively).} \label{fig:NoccO1s}  
\end{figure}  

Further insight into the influence of the contact layer on the structural and electronic properties of  LAO/STO(001)  can be obtained from Fig.~\ref{fig:NoccO1s}. The buckling within the layers expressed as a relative shift of anion and cation $z$-coordinate in Fig.~\ref{fig:NoccO1s} differs significantly  compared to the systems without a metallic contact. As discussed in Section 2, in the uncovered LAO films on STO(001) the lattice polarization is confined to the LAO film where it plays a decisive role to counteract the divergence of the electric potential, while the buckling is  negligible within the STO substrate (see orange lines). In contrast, after adsorbing the metallic overlayer, the potential build up in LAO is largely canceled and hence the anion-cation buckling is strongly reduced in size (Pt contact) and even changes sign (Al and Ti overlayer). 

Surprisingly at first sight, a notable polarization emerges within STO that is strongest at the interface and  decays in deeper layers. The pattern is  similar to the one in LTO/STO and LAO/STO superlattices with an $n$-type interface~\cite{Okamoto:2006,Pentcheva:2008}. This lattice polarization in STO correlates directly with the occupation of Ti $3d$ bands (Fig.~\ref{fig:NoccO1s}b), which is highest in the interface TiO$_2$ layer and decays in deeper layers. Concerning the chemical effects of the overlayer, the lattice polarization and band occupation at the interface are highest  for the system with an Al overlayer, followed by Ti and lowest for a Pt overlayer.
The position of O$1s$ level, which is a monitor of the local potential, shows a similar trend with the strongest binding energy in the interface TiO$_2$ layer for an Al contact layer. The strength in binding to the surface is reflected also in the M-O bond lengths, Al-O being the shortest (1.97~\AA), followed by Ti-O (2.00~\AA) and Pt-O being longest (2.31~\AA) (see Fig.~\ref{fig:structAlTiPt}).
 
{\sl Dependence on the metallic contact thickness} To investigate the dependence on the metallic contact thickness, we have varied the thickness of the Ti overlayer. The calculations for 2ML Ti/2LAO/STO(001) show a similar occupation of the Ti $3d$ band at the interface but a stronger band bending in the deeper layers leading to a lower total number of electrons within STO. This correlates with a slightly larger distance between Ti in the contact layer and oxygen in the top AlO$_2$ layer of 2.00\AA\ vs. 2.06\AA\ for  1ML and 2ML Ti/2LAO/STO(001), respectively. 

\begin{table}
\caption{Calculated $M$-O bond lengths, Ti-O buckling (difference in $z$-coordinate between cations and oxygen) and Ti $3d$ band  occupation in the interface layer, integrated between \ef-0.65 eV and \ef.  Work functions and Schottky barrier heights for the different contact overlayers are also displayed. 
}\label{table1}
\begin{tabular}{|c|c|c|c|c|c|}
\hline 
Metallic  & O-M distance & $z_{\rm Ti}-z_{\rm O}$ (IF)&  $n_{\rm occ}$ & $\Phi$ & $p$-SBH    \\
Layer	& (\AA) & (\AA) &  &(eV) & (eV)        \\
\hline
Al	& 1.97 & -0.20 & 0.15 & 3.53 & 3.0  \\
Ti	& 2.00 & -0.19 & 0.13 & 4.05 & 2.8 \\
2Ti & 2.06 & -0.18 & 0.11 & 4.05 & 2.8  \\
Pt	& 2.31 & -0.08 & 0.04 & 5.60 & 2.2\\
\hline
\end{tabular}
\end{table}

{\sl Schottky barriers} Another illuminating characteristic of the $M$/LAO/STO(001) system, important for understanding the variation of band lineups and in view of device applications, are the Schottky barrier heights. As displayed in Table~\ref{table1} these correlate with the strength of the chemical bond and, most importantly, with the work function: Al has the highest $p$-SBH (3.0~eV), followed by Ti (2.8~eV) and finally Pt (2.2~eV). As the well-known underestimation of band gaps within GGA is mostly related to the position of the conduction band minimum, $n$-SBH tend to be too small. Taking the experimental band gap of LAO, the $n$-SBH are 2.6~eV (Al), 2.8~eV (Ti) and 3.4~eV (Pt). The relevance of the chemical bonding between $M$ and LAO/STO(001) is in line with previous findings for metal-oxide interfaces~\cite{Stengel:naturemat:2009}.

\begin{figure}[t]
 \hspace{-2cm}
   \begin{center}
   \includegraphics[scale=0.5,angle=0]{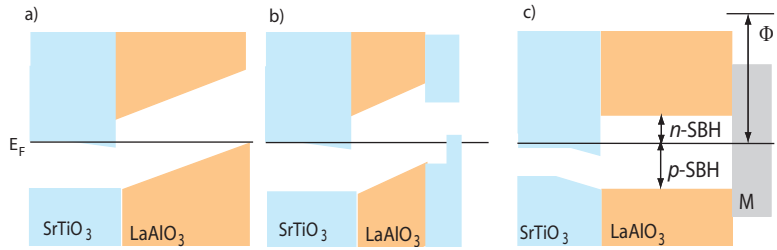}
   \end{center}
\caption{Schematic band diagram of a) LAO/STO(001) at the verge of an electronic reconstruction at the critical LAO thickness; LAO/STO(001) covered by b) an STO overlayer which leads to IMT already at 2 ML LAO and c) a metallic  contact layer ($M$). Note, that the potential build up that leads to an electronic reconstruction in a) and b) beyond a  critical LAO thickness is  eliminated in $M$/LAO/STO(001).}\label{fig:banddiagr} 
\end{figure}
Fig.~\ref{fig:banddiagr} displays the schematic band diagrams of  three distinct mechanisms of formation of a 2DEG in LAO/STO(001), STO/LAO/STO(001) and M/LAO/STO(001). For LAO/STO(001) a thickness dependent IMT occurs as a result of the potential buildup within the LAO film, where the electronic reconstruction involves the formation of holes at the surface and electrons at the interface.  In STO/LAO/STO(001) due to an additional surface state in the top TiO$_2$ layer the IMT takes place at a much lower LAO critical thickness leading to the formation of an electron-hole bilayer.  In contrast, for $M$/LAO/STO(001) ($M$=Al, Ti) the potential in LAO is flat regardless of the LAO or STO thickness. Simultaneously, a 2DEG with higher carrier density is formed at the interface. In the case of Pt (and other transititon and noble metals, for more details see Ref.~\cite{Ruiz2011}), likely due to the weaker bonding and smaller charge transfer to the oxide layer, a residual slope within LAO is found, consistent with the recently measured potential  build up in Pt/LAO/STO(001)~\cite{Bhalla:2011}.

The evolution of the system due to deposition of the contact can be considered in several steps. Beginning with the internal field in 2ML LAO/STO(001) [$\sim$0.8 eV potential rise], the metal layer is added and allowed to equilibrate with the outer LAO layer, determining the work function and the Schottky barrier (contact Fermi level with respect to the LAO band edges). Based on the final result shown in Fig.~\ref{fig:DOSAlTiPt2LAOSTO}, the contact Fermi level would be $\sim$1 eV above the STO conduction band minimum. Now, allowing exchange of charge with the interface -- full self-consistency -- electrons will flow from the contact layer to the interface until the Fermi levels coincide. As charge is transferred,  the internal field and accompanying ionic polarization decrease. The determining factor is the lineup of the surface and interface Fermi levels which is accompanied 
by the (near-)vanishing of the electric field within the LAO slab. This charge transfer between the metal and the surface AlO$_2$ layer, leaving a slightly positive contact layer, will renormalize the work function of the metal and Schottky barrier somewhat. The self-consistent calculation gives only this final result, shown in Fig.~\ref{fig:DOSAlTiPt2LAOSTO}.

	\subsection{Thin \sto{} substrate: Spin polarization}
While the results for $M$/LAO/STO(001) presented above were obtained using a 6.5 ML STO substrate, we have also performed calculations with a thin (2.5 ML) STO and observe there interesting confinement effects that will be described in the following. As shown in Fig.~\ref{fig:NoccO1s}b a much higher Ti $3d$ band occupation is obtained at the interface for the thin STO substrate. However, the most prominent effect is a spin polarization of the 2DEG at the interface for the thin STO substrate. When adsorbing a single Ti layer, Ti displays a significant magnetic moment of 0.60\muB\ due to the reduced coordination of the metal atom at the surface. We note that also a single Pt overlayer is spin-polarized with 0.49\muB\ due to polarization of the holes in the $5d$ shell. For a Ti overlayer, this effect induces a moment of  0.10 (0.24) \muB\ at the Ti sites in IF (IF-1) layer. The values obtained within GGA+\textit{U}, where  effects of strong intra-atomic interaction are included, are enhanced:  0.20 (0.30)\muB\ in IF (IF-1). Such polarization seems surprising since the interface and surface layers are separated by the insulating  LAO slab, precluding (or vastly reducing) exchange coupling between the layers but indeed the effect depends in the LAO spacer thickness: An increase of the width of the LAO layer reduces the influence of the contact layer: for a 4 ML thick LAO film the spin polarization of the 2DEG decreases to 0.05/0.11\muB\ for IF/IF-1. 

This spin-polarization is a result of two effects.  First, the fact that spin-up and spin-down Fermi levels at the surface and interface must be separately aligned (as well as with each other, finally).  Second, a quantum confinement effect creates discrete bands and a classical confinement effect leads to enhanced charge occupation at the IF and IF-1 layers.  The magnetic moment of Ti in the surface contact layer is approximately 0.6\muB\ but decreases once the contact thickness is increased to 2 ML: the magnetic moment of the surface and subsurface Ti layer is reduced to 0.25 \muB\ and -0.05 \muB\ respectively  due to the enhanced Ti coordination.  For other electrodes with lower surface magnetic moment, like for example 1 ML of Al (no magnetic moment) or 2 ML of Ti, the spin-polarization of the 2DEG is quenched.

\section{\label{Broader_Functionalities}Broader functionalities of these interfacial systems}
The range of unexpected phenomena that has been observed in these nanostructures has systematically been broadened by new discoveries.  In the preceding sections we have discussed how unusual properties, such as two-type carrier conduction or magnetism, may arise in these nanostructures. These theoretical studies are based, with a few exceptions, on atomically abrupt, structurally ideal interfaces. It is understood that most if not all samples are more complicated than this. We provide here an overview of the discoveries of magnetism at the interface, then of the reports of superconductivity, and finally of the recent reports of coexistence of these two distinct, and usually strongly competing, types of long-range order. Understanding these phenomena will shed light on the underlying electronic structure at and near the interface.  Specifically, several of these studies point out the importance of inhomogeneities in the samples, complications that should be kept in mind and will finally have to be taken into account.  From the theorist's viewpoint, this emphasizes the importance of controlling defect concentration and distribution, to allow a closer connection of theory to experiment.

\subsection{Hysteresis reflecting magnetic order}
Soon after the initial reports on conductivity in the LAO/STO system  raised interest and research activity on this system, Brinkman and collaborators reported magnetic hysteresis in transport properties at the LAO/STO interface of PLD-deposited films.~\cite{Brinkman:Huijben:etal2007} This hysteresis, and an associated large negative magnetoresistance, 
reflects magnetic order arising at a few Kelvins and moves this oxide interface toward spintronics applications. Even prior to this study the prospect of magnetism at this interface had been raised by Pentcheva and Pickett~\cite{Pentcheva:Pickett:2006} in theoretical studies of LAO/STO superlattices with either $n$ or $p$-type interfaces. The question they faced at the time was this: at the polar LAO/STO interface, there is 0.5 carrier per interface cell too many ($n$-type) or too few ($p$-type) to fill bands. Therefore the interface will be conducting unless other considerations come into play. For a $n$-type interface it was demonstrated that correlation effects within GGA+$U$ give rise to a charge and orbitally ordered state with magnetic moments of 0.7\muB\ at the \tith\ -sites. Later Zhong and Kelly~\cite{Zhong:2008} investigated the effects of structural distortion beyond tetragonal symmetry and its impact on charge and spin ordering, finding an antiferromagnetic ground
  state. The relationship of the observed magnetism (due to Ti$^{3+}$ local moments at the $n$-type interface) to the conduction by carriers remains to be clarified. We note that samples grown at high oxygen pressure, where the effect of oxygen vacancies is minimized, are nearly insulating with a sheet resistance that is seven orders of magnitude higher~\cite{Brinkman:Huijben:etal2007} than the initial samples studied by Ohtomo and Hwang~\cite{Ohtomo:Hwang2004}.
On the other hand, the $p$-type interface had been found {\it always} to be non-conducting, independent of growth parameters.  For a defect-free interface suggested by the initial reports, Pentcheva and Pickett proposed  that correlation effects on the O $2p$ orbitals, resulting in charge order of oxygen holes with a magnetic moment 0.68\muB, provided a plausible mechanism for insulating behavior. While this is an interesting case  of $d^0$-magnetism, the insulating behavior of the $p$-type interface has been generally  ascribed to O vacancies~\cite{Nakagawa:2006}.

The possibility of magnetism at these interfaces was highlighted in Physics Today in 2007~\cite{Goss_Levi:2007} and remains an enigma. Magnetism was further discussed by Huijben {\it et al.},~\cite{Huijben:2009} who provided a broader overview of work on oxide interfaces up to that time.  Subsequently, Seri and Klein~\cite{seri2009} reported antisymmetric magnetoresistance at this interface, and noted magnetic field induced inhomogeneous magnetism, but did not characterize it as spontaneous ferromagnetism. Ariando and collaborators~\cite{ariando2011} detected dia-, ferro- and paramagnetic signals interpreted in terms of phase separation. In particular, the ferromagnetic phase persisted beyond room temperature. Li {\it et al.}~\cite{Li:2011} reported  magnetic effects persisting up to 200 K which raises new questions, beginning with how such apparently weak magnetism can maintain spin order to such high temperature.

\subsection{Superconductivity}
The observation of superconductivity at the $n$-type interface by Reyren {\it et al.}~\cite{Reyren:Thiel:Mannhart2007} further enriched the variety of phenomena that have been reported. The superconducting state in PLD-grown samples at intermediate pressures (10$^{-4}$~mbar) arose below 200 mK. Since bulk $n$-type STO superconducts in that range (up to 400 mK) one might initially question how new the phenomenon really is. The magnetic field directional dependence established that the superconductivity was highly two-dimensional, so in that manner its behavior is quite unlike that of bulk STO; as a 2D system it is characterized by a 2D carrier concentration rather than a 3D one.  As with the magnetism, the degree of localization around the interface, and origin and character of the carriers, have become objects of scrutiny.

Subsequent reports by Caviglia \etal~\cite{caviglia2008} and Ben Shalom \etal~\cite{shalom2010} confirmed superconductivity arising at this interfacial system.
The latter report where 15  unit cells of LAO were deposited with pulsed laser and the carrier density was varied with gating, mapped out a portion of the phase diagram where T$_c$ decreases as the carrier density is increasing. Kim and collaborators~\cite{kim2011} have studied by Shubnikov-de Haas oscillations the (presumably) related superconducting state in $\delta$-doped STO films deposited by PLD. The doping by Nb (for Ti) was performed in a varying number of TiO$_2$ layers, and a 2D to 3D crossover in the superconducting state was monitored. This still very new topic of oxide interface superconductivity has been reviewed by Gariglio and Triscone.~\cite{gariglio2011}

\subsection{Coexistence}
Historically magnetism and superconductivity  were mutual anathema, but the high temperature superconductivity discovered in cuprates changed that. The latest big discovery in superconductivity, the Fe-based pnictides and chalcogenides, also display magnetic phases next to superconducting ones.  These as well as some other materials classes such as heavy fermion metals,~\cite{Sarrao:2002} raise questions about possible connections between these ordered phases. In these examples the magnetism is antiferromagnetism (or correlations). Superconductivity coexisting with {\it ferromagnetism} is very rare, having been reported only in the past decade or so in RuSr$_2$GdCu$_2$O$_8$~\cite{bernhard1999} and in the three uranium compounds UGe$_2$, URhGe, and UCoGe. Aoki {\it et al.} have provided a recent overview summarizing several of the phenomena and some of the issues to account for observations.\cite{aoki2011}  There are several challenges to be overcome for superconductivity 
 to arise in a ferromagnet (see discussion in Ref.~\cite{pickett1999}): exchange splitting of up and down bands that disrupts pairing, magnetic field disruption of superconducting order, etc. We return briefly below to some of these issues.

The first observation of coexistence, by Dikin \etal,~\cite{Dikin:2011} involved superconducting onsets around T$_c\approx $150 mK on pulsed laser deposited (PLD) films of ten unit cells of LAO on TiO$_2$-terminated STO(001) substrates. They observed hysteresis in $T_c$($H$) curves ($H$ is the magnetic field) attributed to underlying ferromagnetic order. Their interpretation focuses on two parallel conduction channels:  One consists of localized, magnetic states that are intrinsic to the interface (viz. Ti$^{3+}$ moments, likely in the immediate interface region), while the other carriers are itinerant electrons contributed by defects (a likely candidate being oxygen vacancies). While previous reports indicated quenching of magnetism at a few Kelvin, Li {\it et al.}~\cite{Li:2011} report high resolution magnetic torque magnetometry measurements, showing evidence of magnetic (``superparamagnetic-like'') order up to 200 K. The superconducting signal in their samples arises below 120 mK. 

Most recently  Bert \etal~\cite{Bert:2011} provided a real space picture of their PLD grown films. They performed imaging with a scanning SQUID (superconducting quantum interference device) over $\sim$200 $\mu$m square regions on samples of 10 ML LAO on STO(001) substrates. They report strong inhomogeneity, with sub-micron regions of ferromagnetism in a background  of paramagnetic carriers that show a diamagnetic superconducting signal around 100 mK: Thus superconductivity and magnetism coexist in the sample but not in the same region. For comparison, no magnetic dipoles were observed in a reference sample of $\delta$-doped STO. From their transport and thermodynamic measurements, they infer that most of the intrinsic carriers (4$\times$10$^{14}$ cm$^{-2}$ per interface cell for an atomically perfect interface) are localized, since an order-of-magnitude fewer carriers contribute to the Hall conductivity.

Some early attempts to account for this coexistence are appearing.~\cite{michaeli2011, stephanos2011,pavlenko2011}. The coexistence seems an even more delicate question here than in the ferromagnetic superconductors mentioned above.  The $\sim$200 mK critical temperature reflects a BCS gap of a few $\mu$eV, a truly tiny energy scale which appreciable exchange splitting of the bands would overwhelm. The critical magnetic field should be very small. There are aspects of the interfaces that suggest means to avoid these pair-breakers. The magnetic electrons and the superconducting carriers may reside in different bands, requiring a generalization of models applied to the previously  known ferromagnetic superconductors. Inhomogeneity or phase separation may play a role, and coexistence in the same sample may not imply local coexistence (in the same region of the sample). The superconducting state may be inhomogeneous, of the FFLO type.\cite{FF,LO}  

\section{Summary}

The \lao/\sto\ system shows a remarkable spectrum of electronic phenomena, some of which seem to be understood and others that require further study. We have provided an overview based on DFT results addressing how the electrostatic boundary conditions determine charge (re-)distribution and the electronic state. A variety  of parameters are identified, such as the surface and interface termination and stoichiometry, the presence of metallic or oxide overlayers, as well as defects and adsorbates, that enable  tuning the electronic behavior of this system. Understanding and
controlling these parameters, and especially defects, dopands~\cite{Blamire:2009} and adsorbates, remains a challenge that needs further attention  in future theoretical and experimental studies. 

\ack{We acknowledge financial support through the DFG SFB/TR80 (project C3)
and  grant {\sl h0721} for computational time at the Leibniz Rechenzentrum. V. G. R. acknowledges financial support from CONACYT (Mexico) and DAAD (Germany). W. E. P. was supported by U.S. Department of Energy Grant No. DE-FG02-04ER46111.}


\begin{thebibliography}{80}
\providecommand{\url}[1]{\texttt{#1}}
\providecommand{\urlprefix}{URL }
\expandafter\ifx\csname urlstyle\endcsname\relax
  \providecommand{\doi}[1]{DOI: \discretionary{}{}{}#1}\else
  \providecommand{\doi}{DOI: \discretionary{}{}{}\begingroup
  \urlstyle{rm}\Url}\fi


\bibitem{Ohtomo:Hwang2004}
  Ohtomo, A. and Hwang, H.Y. 2004 A high-mobility electron gas at the
  {L}a{A}l{O}$_3$/{S}r{T}i{O}$_3$ heterointerface.
 \emph{Nature}, \textbf{427}, 423.
 \doi{10.1038/nature02308}.

\bibitem{Reyren:Thiel:Mannhart2007}
Reyren, N., Thiel, S., Caviglia, A.D., Kourkoutis, L.F., Hammerl, G., Richter,
  C., Schneider, C.W., Kopp, T., R\"uetschi, A.S., Jaccard, D., et~al. 2007
  Superconducting interfaces between insulating oxides.
 \emph{Science}, \textbf{317(5842)}, 1196--1199.
 \doi{10.1126/science.1146006}.

\bibitem{Brinkman:Huijben:etal2007}
Brinkman, A., Huijben, M., van Zalk, M., Huijben, J., Zeitler, U., Maan, J.C.,
  van~der Wiel, W.G., Rijnders, G., Blank, D.H.A., and Hilgenkamp, H. 2007
  Magnetic effects at the interface between non-magnetic oxides.
 \emph{Nature Mater.}, \textbf{6(7)}, 493--496.
 \doi{10.1038/nmat1931}.

\bibitem{Bert:2011}
Bert, J.A., Kalisky, B., Bell, C., Kim, M., Hikita, Y., Hwang, H.Y., and Moler,
  K.A. 2011 Direct imaging of the coexistence of ferromagnetism and
  superconductivity at the {L}a{A}l{O}$_3$/{S}r{T}i{O}$_3$ interface.
 \emph{Nature Physics}, \textbf{7}, 767--771.
 \doi{10.1038/nphys2079}.

\bibitem{Dikin:2011}
Dikin, D.A., Mehta, M., Bark, C.W., Folkman, C.M., Eom, C.B., and
  Chandrasekhar, V. 2011 Coexistence of superconductivity and ferromagnetism in
  two dimensions.
 \emph{Phys. Rev. Lett.}, \textbf{107(5)}, 056802.
 \doi{10.1103/PhysRevLett.107.056802}.

\bibitem{Li:2011}
Li, L., Richter, C., Mannhart, J., and Ashoori, R.C. 2011 Coexistence of
  magnetic order and two-dimensional superconductivity at
  {L}a{A}l{O}$_3$/{S}r{T}i{O}$_3$ interfaces.
 \emph{Nature physics}, \textbf{7}, 762–766.
 \doi{10.1038/nphys2080}.

\bibitem{Thiel:Hammerl:Mannhart2006}
Thiel, S., Hammerl, G., Schmehl, A., Schneider, C.W., and Mannhart, J. 2006
  Tunable quasi-two-dimensional electron gases in oxide heterostructures.
 \emph{Science}, \textbf{313(5795)}, 1942--1945.
 \doi{10.1126/science.1131091}.

\bibitem{Cen:Thiel:Hammerl:etal2008}
Cen, C., Thiel, S., Hammerl, G., Schneider, C.W., Andersen, K.E., Hellberg,
  C.S., Mannhart, J., and Levy, J. 2008 Nanoscale control of an interfacial
  metal-insulator transition at room temperature.
 \emph{Nature Mater.}, \textbf{7(4)}, 298--302.
 \doi{10.1038/nmat2136}.

\bibitem{Bi:Levy:2010}
Bi, F., Bogorin, D.F., Cen, C., Bark, C.W., Park, J.W., Eom, C.B., and Levy, J.
  2010 \textquotedblleft{}{W}ater-cycle\textquotedblright{} mechanism for
  writing and erasing nanostructures at the {L}a{A}l{O}$_3$/{S}r{T}i{O}$_3$
  interface.
 \emph{Appl. Phys. Lett.}, \textbf{97}, 173110.
 \doi{10.1063/1.3506509}.

\bibitem{Chen:2010}
Chen, Y.Z., Zhao, J.L., Sun, J.R., Pryds, N., and Shen, B.G. 2010 Resistance
  switching at the interface of {L}a{A}l{O}$_3$/{S}r{T}i{O}$_3$.
 \emph{Appl. Phys. Lett.}, \textbf{97}, 123102.
 \doi{10.1063/1.3490646}.

\bibitem{Cen:2009}
Cen, C., Thiel, S., Mannhart, J., and Levy, J. 2009 Oxide nanoelectronics on
  demand.
 \emph{Science}, \textbf{323(5917)}, 1026--1030.
 \doi{10.1126/science.1168294}.

\bibitem{Bogorin:2010}
Bogorin, D.F., Irvin, P., Cen, C., and Levy, J. 2010
  La{A}l{O}$_3$/{S}r{T}i{O}$_3$-{B}ased device concepts.
 \emph{arXiv:1011.5290v1}.

\bibitem{PentchevaPRL:2010}
Pentcheva, R., Huijben, M., Otte, K., Pickett, W.E., Kleibeuker, J.E., Huijben,
  J., Boschker, H., Kockmann, D., Siemons, W., Koster, G., et~al. 2010 Parallel
  electron-hole bilayer conductivity from electronic interface reconstruction.
 \emph{Phys. Rev. Lett.}, \textbf{104(16)}, 166804.
 \doi{10.1103/PhysRevLett.104.166804}.

\bibitem{Hwang2006}
Hwang, H.Y. 2006 Tuning interface states.
 \emph{Science}, \textbf{313(5795)}, 1895--1896.
 \doi{10.1126/science.1133138}.

\bibitem{Goniakowski:2008}
Goniakowski, J., Finocchi, F. and Noguera, C. 2008 Polarity of oxide surfaces and
  nanostructures.
 \emph{Rep. Prog. Phys.}, \textbf{71(1)}, 016501.
 \doi{10.1088/0034-4885/71/1/016501}.

\bibitem{Pauli:2008}
Pauli, S.A. and Willmott, P.R. 2008 Conducting interfaces between polar and
  non-polar insulating perovskites.
 \emph{J. Phys.: Condens. Matter}, \textbf{20(26)}, 264012.
 \doi{10.1088/0953-8984/20/26/264012}.

\bibitem{Huijben:2009}
Huijben, M., Brinkman, A., Koster, G., Rijnders, G., Hilgenkamp, H., and Blank,
  D.H.A. 2009 Structure-property relation of {S}r{T}i{O}$_3$/{L}a{A}l{O}$_3$
  interfaces.
 \emph{Adv. Mater.}, \textbf{21(17)}, 1665--1677.
 \doi{10.1002/adma.200801448}.

\bibitem{Pentcheva:2010}
Pentcheva, R. and Pickett, W.E. 2010 Electronic phenomena at complex oxide
  interfaces: insights from first principles.
 \emph{J. Phys.: Condens. Matter}, \textbf{22}, 043001.
 \doi{10.1088/0953-8984/22/4/043001}.

\bibitem{Chen:Kolpak:Ismail2010}
Chen, H., Kolpak, A.M., and {Ismail-Beigi}, S. 2010 Electronic and magnetic
  properties of {S}r{T}i{O}$_3$/{L}a{A}l{O}$_3$ interfaces from first
  principles.
 \emph{Adv. Mater.}, \textbf{22(26-27)}, 2881--2899.
 \doi{10.1002/adma.200903800}.

\bibitem{Triscone:2011}
Zubko, P., Gariglio, S., Gabay, M., Ghosez, P., and Triscone, J.M. 2011
  Interface physics in complex oxide heterostructures.
 \emph{Annu. Rev. Condens. Matter Phys.}, \textbf{2}, 141--165.
 \doi{10.1146/annurev-conmatphys-062910-140445}.

\bibitem{Stengel:2011}
Stengel, M. 2011 First-principles modeling of electrostatically doped
  perovskite systems.
 \emph{Phys. Rev. Lett.}, \textbf{106}, 136803.
 \doi{10.1103/PhysRevLett.106.136803}.

\bibitem{Xie:Hwang:2010}
Xie, Y., Bell, C., Yajima, T., Hikita, Y., and Hwang, H.Y. 2010 Charge writing
  at the {L}a{A}l{O}$_3$/{S}r{T}i{O}$_3$ surface.
 \emph{Nano Lett.}, \textbf{10(7)}, 2588--2591.
 \doi{10.1021/nl1012695}.

\bibitem{Segal:Reiner:etal2009}
Segal, Y., Ngai, J.H., Reiner, J.W., Walker, F.J., and Ahn, C.H. 2009 X-ray
  photoemission studies of the metal-insulator transition in
  {L}a{A}l{O}$_{3}$/{S}r{T}i{O}$_{3}$ structures grown by molecular beam
  epitaxy.
 \emph{Phys. Rev. B}, \textbf{80(24)}, 241107.
 \doi{10.1103/PhysRevB.80.241107}.

\bibitem{Sing:Claessen:2009}
Sing, M., Berner, G., Go\ss{}, K., M\"uller, A., Ruff, A., Wetscherek, A.,
  Thiel, S., Mannhart, J., Pauli, S.A., Schneider, C.W., et~al. 2009 Profiling
  the interface electron gas of {L}a{A}l{O}$_{3}$/{S}r{T}i{O}$_{3}$
  heterostructures with hard {X}-ray photoelectron spectroscopy.
 \emph{Phys. Rev. Lett.}, \textbf{102(17)}, 176805.
 \doi{10.1103/PhysRevLett.102.176805}.

\bibitem{Chambers:Ramasse:2010}
Chambers, S., Engelhard, M., Shutthanandan, V., Zhu, Z., Droubay, T., Qiao, L.,
  Sushko, P., Feng, T., Lee, H., Gustafsson, T., et~al. 2010 Instability,
  intermixing and electronic structure at the epitaxial heterojunction.
 \emph{Surface Science Reports}, \textbf{65(10-12)}, 317 -- 352.
 \doi{10.1016/j.surfrep.2010.09.001}.

\bibitem{Zhong:2010}
Zhong, Z., Xu, P.X., and Kelly, P.J. 2010 Polarity-induced oxygen vacancies at
  {L}a{A}l{O}$_{3}$/{S}r{T}i{O}$_{3}$ interfaces.
 \emph{Phys. Rev. B}, \textbf{82(16)}, 165127.
 \doi{10.1103/PhysRevB.82.165127}.

\bibitem{Bristowe:2011}
Bristowe, N.C., Littlewood, P.B., and Artacho, E. 2011 Surface defects and
  conduction in polar oxide heterostructures.
 \emph{Phys. Rev. B}, \textbf{83(20)}, 205405.
 \doi{10.1103/PhysRevB.83.205405}.

\bibitem{Son:2010}
Son, W., Cho, E., Lee, J., and Han, S. 2010 Hydrogen adsorption and carrier
  generation in {L}a{A}l{O}$_3$-{S}r{T}i{O}$_3$ heterointerfaces: {A}
  first-principles study.
 \emph{J. Phys.: Condens. Matter}, \textbf{22(31)}, 315501.
 \doi{10.1088/0953-8984/22/31/315501}.

\bibitem{Willmott:2007}
Willmott, P.R., Pauli, S.A., Herger, R., Schlep\"utz, C.M., Martoccia, D.,
  Patterson, B.D., Delley, B., Clarke, R., Kumah, D., Cionca, C., et~al. 2007
  Structural basis for the conducting interface between {L}a{A}l{O}$_{3}$ and
  {S}r{T}i{O}$_{3}$.
 \emph{Phys. Rev. Lett.}, \textbf{99(15)}, 155502.
 \doi{10.1103/PhysRevLett.99.155502}.

\bibitem{Qiao:2010}
Qiao, L., Droubay, T.C., Shutthanandan, V., Zhu, Z., Sushko, P.V., and
  Chambers, S.A. 2010 Thermodynamic instability at the stoichiometric
  {L}a{A}l{O}$_3$/{S}r{T}i{O}$_3$(001) interface.
 \emph{J. Phys.: Condens. Matter}, \textbf{22}, 312201.
 \doi{10.1088/0953-8984/22/31/312201}.

\bibitem{Ishibashi:Terakura2008}
Ishibashi, S. and Terakura, K. 2008 Analysis of screening mechanisms for polar
  discontinuity for {L}a{A}l{O}$_3$/{S}r{T}i{O}$_3$ thin films based on
  \textit{Ab initio} calculations.
 \emph{J. Phys. Soc. Jpn.}, \textbf{77(10)}, 104706.
 \doi{10.1143/JPSJ.77.104706}.

\bibitem{Pentcheva:Pickett2009}
Pentcheva, R. and Pickett, W.E. 2009 Avoiding the polarization catastrophe in
  {L}a{A}l{O}$_{3}$ overlayers on {S}r{T}i{O}$_{3}$(001) through polar
  distortion.
 \emph{Phys. Rev. Lett.}, \textbf{102(10)}, 107602.
 \doi{10.1103/PhysRevLett.102.107602}.

\bibitem{Son:2009}
Son, W.j., Cho, E., Lee, B., Lee, J., and Han, S. 2009 Density and spatial
  distribution of charge carriers in the intrinsic $n$-type
  {L}a{A}l{O}$_{3}$--{S}r{T}i{O}$_{3}$ interface.
 \emph{Phys. Rev. B}, \textbf{79(24)}, 245411.
 \doi{10.1103/PhysRevB.79.245411}.

\bibitem{Pauli:2011}
Pauli, S.A., Leake, S.J., Delley, B., Bj\"orck, M., Schneider, C.W.,
  Schlep\"utz, C.M., Martoccia, D., Paetel, S., Mannhart, J., and Willmott,
  P.R. 2011 Evolution of the interfacial structure of {L}a{A}l{O}$_3$ on
  {S}r{T}i{O}$_3$.
 \emph{Phys. Rev. Lett.}, \textbf{106(3)}, 036101.
 \doi{10.1103/PhysRevLett.106.036101}.

\bibitem{Nakagawa:2006}
Nakagawa, N., Hwang, H.Y., and Muller, D.A. 2006 Why some interfaces cannot be
  sharp.
 \emph{Nature Mater.}, \textbf{5}, 204--209.
 \doi{10.1038/nmat1569}.

\bibitem{Pavlenko:2011}
Pavlenko, N. and Kopp, T. 2004 Structural relaxation and metal-insulator
  transition at the interface between {S}r{T}i{O}$_{3}$ and {L}a{A}l{O}$_3$.
 \emph{Surface Science}, \textbf{605}, 1114--1121.
 \doi{10.1016/j.susc.2011.03.016}.

\bibitem{Huijben:2006}
Huijben, M., Rijnders, G., Blank, D.H.A., Bals, S., {Van Aert}, S., Verbeeck,
  J., {Van Tendeloo}, G., Brinkman, A., and Hilgenkamp, H. 2006 Electronically
  coupled complementary interfaces between perovskite band insulators.
 \emph{Nature Mater.}, \textbf{5(7)}, 556--560.
 \doi{10.1038/nmat1675}.

\bibitem{Padilla1998}
Padilla, J. and Vanderbilt, D. 1998 Ab initio study of {S}r{T}i{O}$_{3}$
  surfaces.
 \emph{Surface Science}, \textbf{418(1)}, 64--70.
 \doi{10.1016/S0039-6028(98)00670-0}.

\bibitem{Kimura1995}
Kimura, S., Yamauchi, J., Tsukada, M., and Watanabe, S. 1995 First-principles
  study on electronic structure of the (001) surface of {S}r{T}i{O}$_3$.
 \emph{Phys. Rev. B}, \textbf{51(16)}, 11049--11054.
 \doi{10.1103/PhysRevB.51.11049}.

\bibitem{Hwangcondmat}
Takizawa, M., Tsuda, S., Susaki, T., Hwang, H.Y., and Fujimori, A. 2011
  Electronic charges and electric potential at {L}a{A}l{O}$_3$/{S}r{T}i{O}$_3$
  interfaces studied by core-level photoemission spectroscopy.
 \emph{Phys. Rev. B}, \textbf{84}, 245124.
 \doi{10.1103/PhysRevB.84.245124}.

\bibitem{ChenPRB:2009}
Chen, H., Kolpak, A.M., and Ismail-Beigi, S. 2009 Fundamental asymmetry in
  interfacial electronic reconstruction between insulating oxides: An
  \textit{ab initio} study.
 \emph{Phys. Rev. B}, \textbf{79}, 161402.
 \doi{10.1103/PhysRevB.79.161402}.

\bibitem{Zhang:2010}
Zhang, L., Zhou, X.F., Wang, H.T., Xu, J.J., Li, J., Wang, E.G., and Wei, S.H.
  2010 Origin of insulating behavior of the $p$-type
  {L}a{A}l{O}$_3$/{S}r{T}i{O}$_3$ interface: {P}olarization-induced asymmetric
  distribution of oxygen vacancies.
 \emph{Phys. Rev. B}, \textbf{82(12)}, 125412.
 \doi{10.1103/PhysRevB.82.125412}.

\bibitem{Li:2009}
Li, Y., {Na Phattalung}, S., Limpijumnong, S., Kim, J., and Yu, J. 2009
  Formation of oxygen vacancies and charge carriers induced in the $n$-type interface of a LaAlO${}_{3}$ overlayer on SrTiO${}_{3}$(001)
 \emph{Phys. Rev. B}, \textbf{84}, 245307.

\bibitem{pavlenko2011}
Pavlenko, N., Kopp, T., Y.Tsymbal, E., Sawatzky, G.A., and Mannhart, J. 2012
  Magnetism and superconductivity at {LAO}/{STO}-interfaces both generated by
  the {T}i 3d interface electrons?
 \emph{Phys. Rev. B}, \textbf{85(2)}, 020407.
 \doi{10.1103/PhysRevB.85.020407}.

\bibitem{Jany:2010}
Jany, R., Breitschaft, M., Hammerl, G., Horsche, A., Richter, C., Paetel, S.,
  Mannhart, J., Stucki, N., Reyren, R., Gariglio, S., et~al. 2010 Diodes with
  breakdown voltages enhanced by the metal-insulator transition of
  {L}a{A}l{O}$_3$--{S}r{T}i{O}$_3$ interfaces.
 \emph{Appl. Phys. Lett.}, \textbf{96}, 183504.
 \doi{10.1063/1.3428433}.

\bibitem{Bhalla:2011}
Singh-Bhalla, G., Bell, C., Ravichandran, J., Siemons, W., Hikita, Y.,
  Salahuddin, S., Hebard, A.F., Hwang, H.Y., and Ramesh, R. 2011 Built-in and
  induced polarization across {L}a{A}l{O}$_3$/{S}r{T}i{O}$_3$ heterojunctions.
 \emph{Nature Physics}, \textbf{7}, 80.
 \doi{10.1038/NPHYS1814}.

\bibitem{Campbell:1997}
Campbell, C.T. 1997 Ultrathin metal films and particles on oxide surfaces:
  structural, electronic and chemisorptive properties.
 \emph{Surface Science Reports}, \textbf{27(1-3)}, 1--111.
 \doi{10.1016/S0167-5729(96)00011-8}.

\bibitem{Goniakowski:2004}
Goniakowski, J. and Noguera, C. 2004 Electronic states and schottky barrier
  height at metal/{M}g{O}(100) interfaces.
 \emph{Interface science}, \textbf{12(1)}, 93--103.
 \doi{10.1023/B:INTS.0000012298.34540.50}.

\bibitem{Duan:2006}
Duan, C.G., Jaswal, S.S., and Tsymbal, E.Y. 2006 Predicted magnetoelectric
  effect in {F}e/{B}a{T}i{O}$_3$ multilayers: {F}erroelectric control of
  magnetism.
 \emph{Phys. Rev. Lett.}, \textbf{97(4)}, 047201.
 \doi{10.1103/PhysRevLett.97.047201}.

\bibitem{Fu:2007}
Fu, Q. and Wagner, T. 2007 Interaction of nanostructured metal overlayers with
  oxide surfaces.
 \emph{Surface Science Reports}, \textbf{62(11)}, 431--498.
 \doi{10.1016/j.surfrep.2007.07.001}.

\bibitem{Fechner:2008}
Fechner, M., Maznichenko, I.V., Ostanin, S., Ernst, A., Henk, J., Bruno, P.,
  and Mertig, I. 2008 Magnetic phase transition in two-phase multiferroics
  predicted from first principles.
 \emph{Phys. Rev. B}, \textbf{78(21)}, 212406.
 \doi{10.1103/PhysRevB.78.212406}.

\bibitem{Ruiz2011}
Arras, R., and Ruiz, V. G., and Pickett, W. E., and Pentcheva, R. 2012 Tuning the
  two-dimensional electron gas at the {L}a{A}l{O}$_{3}$/{S}r{T}i{O}$_{3}$(001)
  interface by metallic contacts.
 \emph{Phys. Rev. B}, \textbf{85(12)}, 125404.
 \doi{10.1103/PhysRevB.85.125404}.

\bibitem{Wien2k}
Blaha, P., Schwarz, K., Madsen, G., Kvasnicka, D., and Luitz, J.
 {WIEN}2k, an augmented plane wave plus local orbitals program for
  calculating crystal properties.
 (Karlheinz Schwarz, Technische Universit\"at Wien, Austria, 2001).

\bibitem{GGA}
Perdew, J.P., Burke, K., and Ernzerhof, M. 1996 {G}eneralized {G}radient
  {A}pproximation made simple.
 \emph{Phys. Rev. Lett.}, \textbf{77(18)}, 3865--3868.
 \doi{10.1103/PhysRevLett.77.3865}.

\bibitem{Anisimov:1993}
Anisimov, V.I., Solovyev, I.V., Korotin, M.A., Czy\ifmmode~\dot{z}\else
  \.{z}\fi{}yk, M.T., and Sawatzky, G.A. 1993 Density-functional theory and
  {N}i{O} photoemission spectra.
 \emph{Phys. Rev. B}, \textbf{48(23)}, 16929--16934.
 \doi{10.1103/PhysRevB.48.16929}.

\bibitem{Asthagiri:Sholl:2006}
Asthagiri, A. and Sholl, D.S. 2006 Pt thin films on the polar
  {L}a{A}l{O}$_{3}$(100) surface: {A} first-principles study.
 \emph{Phys. Rev. B}, \textbf{73(12)}, 125432.
 \doi{10.1103/PhysRevB.73.125432}.

\bibitem{Asthagiri:Sholl:2002}
Asthagiri, A. and Scholl, D.S. 2002 First principles study of {P}t adhesion and
  growth on {S}r{O}- and {T}i{O}$_{2}$-terminated {S}r{T}i{O}$_{3}$(100).
 \emph{J. Chem. Phys.}, \textbf{116}, 9914--9925.
 \doi{10.1063/1.1476322}.

\bibitem{Ochs:2001}
Ochs, T., K\"ostlmeier, S., and Els\"asser, C. 2001 Microscopic structure and
  bonding at the {P}d/{S}r{T}i{O}$_3$ (001) interface an ab-initio
  local-density-functional study.
 \emph{Integrated Ferroelectrics}, \textbf{32(1-4)}, 267--278.
 \doi{10.1080/10584580108215697}.

\bibitem{Oleinik:2001}
Oleinik, I.I., Tsymbal, E.Y., and Pettifor, D.G. 2001 Atomic and electronic
  structure of {C}o/{S}r{T}i{O}$_{3}$/{C}o magnetic tunnel junctions.
 \emph{Phys. Rev. B}, \textbf{65(2)}, 020401.
 \doi{10.1103/PhysRevB.65.020401}.

\bibitem{Okamoto:2006}
Okamoto, S., Millis, A.J., and Spaldin, N.A. 2006 Lattice relaxation in oxide
  heterostructures: {L}a{T}i{O}$_3$/{S}r{T}i{O}$_3$ superlattices.
 \emph{Phys. Rev. Lett.}, \textbf{97}, 056802.
 \doi{10.1103/PhysRevLett.97.056802}.

\bibitem{Pentcheva:2008}
Pentcheva, R. and Pickett, W.E. 2008 Ionic relaxation contribution to the
  electronic reconstruction at the $n$-type {L}a{A}l{O}$_3$/{S}r{T}i{O}$_3$
  interface.
 \emph{Phys. Rev. B}, \textbf{78}, 205106.
 \doi{10.1103/PhysRevB.78.205106}.

\bibitem{Stengel:naturemat:2009}
Stengel, M., Vanderbilt, D., and Spaldin, N.A. 2009 Enhancement of
  ferroelectricity at metal-oxide interfaces.
 \emph{Nature Mater.}, \textbf{8}, 392--397.
 \doi{10.1038/NMAT2429}.

\bibitem{Pentcheva:Pickett:2006}
Pentcheva, R. and Pickett, W.E. 2006 Charge localization or itineracy at
  {L}a{A}l{O}$_3$/{S}r{T}i{O}$_3$ interfaces: Hole polarons, oxygen vacancies,
  and mobile electrons.
 \emph{Phys. Rev. B}, \textbf{74(3)}, 035112.
 \doi{10.1103/PhysRevB.74.035112}.

\bibitem{Zhong:2008}
Zhong, Z. and Kelly, P. 2008 Electronic-structure-induced reconstruction and
  magnetic ordering at the {L}a{A}l{O}$_3$/{S}r{T}i{O}$_3$ interface.
 \emph{Europhys. Lett.}, \textbf{84}, 27001.
 \doi{10.1209/0295-5075/84/27001}.

\bibitem{Goss_Levi:2007}
{Goss Levi}, B. 2007 Interface between nonmagnetic insulators may be
  ferromagnetic and conducting.
 \emph{Physics Today}, \textbf{60}, 23.
 \doi{10.1063/1.2754590}.

\bibitem{seri2009}
Seri, S. and Klein, L. 2009 Antisymmetric magnetoresistance of the
  {S}r{T}i{O}$_3$/{L}a{A}l{O}$_3$ interface.
 \emph{Phys. Rev. B}, \textbf{80(18)}, 180410.
 \doi{10.1103/PhysRevB.80.180410}.

\bibitem{ariando2011}
Ariando, Wang, X., Baskaran, G., Liu, Z.Q., Huijben, J., Yi, J.B., Annadi, A.,
  Barman, A.R., Rusydi, A., Dhar, S., et~al. 2011 Electronic phase separation
  at the {L}a{A}l{O}$_3$/{S}r{T}i{O}$_3$ interface.
 \emph{Nature Commun.}, \textbf{2}, 188.
 \doi{10.1038/ncomms1192}.

\bibitem{caviglia2008}
Caviglia, A.D., Gariglio, S., Reyren, N., Jaccard, D., Schneider, T., Gabay,
  M., Thiel, S., Hammerl, G., Mannhart, J., and Triscone, J.M. 2008 Electric
  field control of the {L}a{A}l{O}$_3$/{S}r{T}i{O}$_3$ interface ground state.
 \emph{Nature}, \textbf{456}, 624--627.
 \doi{10.1038/nature07576}.

\bibitem{shalom2010}
Ben~Shalom, M., Sachs, M., Rakhmilevitch, D., Palevski, A., and Dagan, Y. 2010
  Tuning spin-orbit coupling and superconductivity at the
  {S}r{T}i{O}$_3$/{L}a{A}l{O}$_3$ interface: A magnetotransport study.
 \emph{Phys. Rev. Lett.}, \textbf{104(12)}, 126802.
 \doi{10.1103/PhysRevLett.104.126802}.

\bibitem{kim2011}
Kim, M., Bell, C., Kozuka, Y., Kurita, M., Hikita, Y., and Hwang, H.Y. 2011
  Fermi surface and superconductivity in low-density high-mobility
  $\delta$-doped {S}r{T}i{O}$_3$.
 \emph{Phys. Rev. Lett.}, \textbf{107(10)}, 106801.
 \doi{10.1103/PhysRevLett.107.106801}.

\bibitem{gariglio2011}
Gariglio, S. and Triscone, J.M. 2011 Oxide interface superconductivity --
  {S}upraconductivit\'e \`a l'interface d'oxydes.
 \emph{Compt. Rend. Physiq.}, \textbf{12(5--6)}, 591--599.
 \doi{10.1016/j.crhy.2011.03.006}.

\bibitem{Sarrao:2002}
Sarrao, J.L., Morales, L.A., Thompson, J.D., Scott, B.L., Stewart, G.R.,
  Wastin, F., Rebizant, J., Boulet, P., Colineau, E., and Lander, G.H. 2002
  Plutonium-based superconductivity with a transition temperature above 18 {K}.
 \emph{Nature}, \textbf{420}, 297--299.
 \doi{10.1038/nature01212}.

\bibitem{bernhard1999}
Bernhard, C., Tallon, J.L., Niedermayer, C., Blasius, T., Golnik, A.,
  Br\"ucher, E., Kremer, R.K., Noakes, D.R., Stronach, C.E., and Ansaldo, E.J.
  1999 Coexistence of ferromagnetism and superconductivity in the hybrid
  ruthenate-cuprate compound {R}u{S}r$_2${G}d{C}u$_2${O}$_8$ studied by muon
  spin rotation and dc magnetization.
 \emph{Phys. Rev. B}, \textbf{59(21)}, 14099--14107.
 \doi{10.1103/PhysRevB.59.14099}.

\bibitem{aoki2011}
Aoki, D., Hardy, F., Miyake, A., Taufour, V., Matsuda, T.D., and Flouquet, J.
  2011 Properties of ferromagnetic superconductors -- {D}es supraconducteurs
  ferromagn\'etiques.
 \emph{Compt. Rend. Physiq.}, \textbf{12(5--6)}, 573--583.
 \doi{10.1016/j.crhy.2011.04.007}.

\bibitem{pickett1999}
Pickett, W.E., Weht, R., and Shick, A.B. 1999 Superconductivity in
  ferromagnetic {R}u{S}r$_2${G}d{C}u$_2${O}$_8$.
 \emph{Phys. Rev. Lett.}, \textbf{83(18)}, 3713--3716.
 \doi{10.1103/PhysRevLett.83.3713}.

\bibitem{michaeli2011}
Michaeli, K., Potter, A.C., and Lee, P.A. 2012 Superconductivity and
  ferromagnetism in oxide interface structures: Possibility of finite momentum
  pairing.
  \emph{Phys. Rev. Lett.}, \textbf{108}, 117003.

\bibitem{stephanos2011}
Stephanos, C., Kopp, T., Mannhart, J., and Hirschfeld, P.J. 2011
  Interface-induced d-wave pairing.
 \emph{Phys. Rev. B}, \textbf{84(10)}, 100510.
 \doi{10.1103/PhysRevB.84.100510}.

\bibitem{FF}
Fulde, P. and Ferrell, R.A. 1964 Superconductivity in a strong spin-exchange
  field.
 \emph{Phys. Rev.}, \textbf{135(3A)}, A550--A563.
 \doi{10.1103/PhysRev.135.A550}.

\bibitem{LO}
Larkin, A.I. and Ovchinnikov, Y.N. 1964 \emph{Zh. Eksp. Teor. Fiz},
  \textbf{47}, 1136--1146.

\bibitem{Blamire:2009}
Fix, T., Schoofs, F., MacManus-Driscoll, J.L., and Blamire, M.G. 2009 Charge
  confinement and doping at {L}a{A}l{O}$_{3}$/{S}r{T}i{O}$_{3}$ interfaces.
 \emph{Phys. Rev. Lett.}, \textbf{103}, 166802.
 \doi{10.1103/PhysRevLett.103.166802}.

\end{thebibliography}

\end{document}